\documentclass[fleqn,usenatbib,useAMS]{mnras}

\usepackage{newtxtext,newtxmath}

\usepackage[T1]{fontenc}
\usepackage{ae,aecompl}

\usepackage{graphicx}	
\usepackage{amsmath}	
\usepackage{amssymb}	
\usepackage[separate-uncertainty = true,multi-part-units=brackets]{siunitx}      
\usepackage{commath}      
\usepackage{physics}  
\usepackage[noabbrev]{cleveref}     

\usepackage[normalem]{ulem}
\usepackage[dvipsnames]{xcolor}

\crefformat{equation}{\eqA equation~\eqB #2#1#3)}
\newcommand{\eqA}{}
\newcommand{\eqB}{(}

\crefname{figure}{Fig.}{Figs.}
\Crefname{figure}{Fig.}{Figs.}
\crefname{table}{Table}{Tables}
\Crefname{equation}{Equation}{Equations}
\newcommand{\solar}{\ensuremath{_{\odot}}} 
\DeclareSIUnit\msol{M\solar{}}
\DeclareSIUnit\lsol{L\solar{}}
\DeclareSIUnit\rsol{R\solar{}}
\DeclareSIUnit\zsol{Z\solar{}}
\DeclareSIUnit\erg{erg}
\DeclareSIUnit\jy{Jy}
\DeclareSIUnit\yr{yr}
\DeclareSIUnit\micron{\micro\metre}
\DeclareSIUnit\au{au}
\DeclareSIUnit\pc{pc}
\DeclareSIUnit\sr{sr}
\DeclareSIUnit\str{sr}
\DeclareSIUnit\kms{\km\per\s}
\newcommand{\mdot}{\ensuremath{\dot{M}}} 

\title[Stellar winds and photoionization in a spiral arm]{Stellar winds and photoionization in a spiral arm}

\author[A. A. Ali, T. J. R. Bending and C. L. Dobbs]{
Ahmad A. Ali\thanks{E-mail: A.Ali2@exeter.ac.uk}, 
Thomas J. R. Bending and
Clare L. Dobbs
\\
Department of Physics and Astronomy, University of Exeter, Stocker Road, Exeter EX4 4QL, United Kingdom
}

\date{Accepted 2021 December 27. Received 2021 December 10; in original form 2021 October 18}

\pubyear{2022}

\begin{document}
\label{firstpage}
\pagerange{\pageref{firstpage}--\pageref{lastpage}}
\maketitle

\begin{abstract}
The role of different stellar feedback mechanisms in giant molecular clouds is not well understood. This is especially true for regions with many interacting clouds as would be found in a galactic spiral arm.
In this paper, building on previous work by Bending et al., we extract a $500\times500\times\SI{100}{\pc}$ section of a spiral arm from a galaxy simulation. We use smoothed particle hydrodynamics (SPH) to re-simulate the region at higher resolution (\SI{1}{\msol} per particle). We present a method for momentum-driven stellar winds from main sequence massive stars, and include this with photoionization, self-gravity, a galactic potential, and ISM heating/cooling. We also include cluster-sink particles with accretion radii of \SI{0.78}{\pc} to track star/cluster formation. The feedback methods are as robust as previous models on individual cloud scales (e.g. Dale et al.).
We find that photoionization dominates the disruption of the spiral arm section, with stellar winds only producing small cavities (at most $\sim$ \SI{30}{\pc}). Stellar winds do not affect the resulting cloud statistics or the integrated star formation rate/efficiency, unlike ionization, which produces more stars, and more clouds of higher density and higher velocity dispersion compared to the control run without feedback. Winds do affect the sink properties, distributing star formation over more low-mass sinks ($\sim$\SI{e2}{\msol}) and producing fewer high-mass sinks ($\sim$\SI{e3}{\msol}). Overall, stellar winds play at best a secondary role compared to photoionization, and on many measures, they have a negligible impact.

\end{abstract}






\section{Introduction}
Star formation takes place in giant molecular clouds (GMCs). Massive stars above \SI{8}{\msol} feed energy and momentum back into GMCs, through processes such as photoionization, stellar winds, radiation pressure, and supernovae (SNe). Further star formation may be induced by feedback through the compression of gas reservoirs, followed by fragmentation \citep{elmegreen1977,whitworth1994}. However, star formation could also be hindered if the feedback processes heat or disperse the gas instead \citep{krumholz2007a,bate2009,walch2012}. Such processes must be understood in order to explain the inefficiency of star formation, wherein only a few per cent of the mass in GMCs is converted into stars \citep{lada2003}. 

Over the lifetime of a massive star ($\sim$3 to \SI{10}{\mega\yr}), the expansion of shells and the flow of turbulent gas will reach length scales beyond the individual cloud size ($\sim$\SI{10}{\pc}), meaning there will be interactions between neighbouring molecular clouds. Furthermore, the GMCs themselves are only component parts of a dynamically evolving galactic environment, as clouds are subject to a global potential and shear which will affect their formation and evolution \citep{dobbs2013a}. 

Until recently, simulations have focused either on individual cloud scales or global galactic scales. The former allows (sub-)parsec resolution of star formation or feedback, and has given insight into gas expulsion \citep[e.g.][]{walch2012,colin2013,rogers2013,ali2018,ali2019,ali2021}, turbulence driving \citep{gritschneder2009,medina2014,sartorio2021}, and the efficiency of star formation \citep{dale2014,geen2018,kim2018a}. However, this almost always involves simplified initial conditions such as spherical clouds, turbulent velocity fields tuned to provide the required boundness, and evolution occurring in isolation without external terms such as gravitational potentials, mass inflow, or radiation fields. 

The opposite is true for simulations of Milky Way-mass galaxies \citep[e.g.][]{agertz2013,dobbs2013a,smith2020,pettitt2020a}, which do model the properties of and interaction between neighbouring GMCs, but rely on subgrid models for star formation and feedback processes in order to run for 100s of Myr. It is usually assumed that SN feedback is the dominant mechanism on galactic scales, allowing pre-SN processes to be neglected. For example, implementations of feedback may involve randomised inputs of energy which are not tied to stellar properties such as mass or lifetime. Furthermore, radiative transfer is computationally expensive and is often neglected on large scales. However, as methods improve, numerical studies on galactic scales are increasingly highlighting the importance of pre-SN feedback in the form of radiation and continuous stellar winds \citep{hopkins2018a}. This is also the indication from observations, e.g. \citet{chevance2020} who inferred GMC dispersal time-scales of a few Myr and lifetimes of the order of \SI{10}{Myr}, by measuring the spatial (de)correlation of clouds and stars in nearby disc galaxies.

Photoionizing feedback has been included in many 3D numerical studies of GMCs over the last decade \citep[e.g.][]{dale2005,mellema2006,peters2010,arthur2011,walch2012,colin2013,geen2015,howard2016,gavagnin2017,ali2018,kim2018a,zamora-aviles2019,vandenbroucke2019a,bending2020,fukushima2020,sartorio2021}. However, fewer studies have focused on stellar winds \citep{dale2008,rogers2013,rey-raposo2017,offner2018,wareing2018}, particularly in combination with radiation \citep{dale2014,ngoumou2015,haid2018,geen2021}. Winds are a difficult problem to solve computationally due to the extreme temperatures (\SI{e7}{K}) and velocities (\SI{e3}{\kms}) involved, as well the radiative processes required to model the cooling of hot, shocked gas (and the spatial resolution needed to resolve this). 

The theoretical model of \citet{weaver1977} describes the interaction of a stellar wind with the ISM, with a free-streaming wind in the innermost region, followed by shocked wind material, then a swept-up shell of shocked ISM gas, finally bounded by the ambient ISM. In this picture, the bubble expands adiabatically as radiative cooling is inefficient for the hot, low-density shocked wind. However, instabilities at the contact discontinuity between the shocked wind and cold, high-density shocked ISM could lead to mixing of the two phases, allowing the shocked wind to cool \citep{capriotti2001}. With efficient cooling, the expansion of the bubble is driven by the ram pressure of the free-streaming wind colliding with the mixed shell. This has been found to occur in simulations of turbulent clouds by \citet{geen2021} and \citet{lancaster2021}. This extreme case permits simpler implementations of feedback in the form of momentum-conserving winds \citep{dale2008}.

Observational measurements have generally inferred the role of stellar winds to be secondary to photoionization. This has been found for regions in the Magellanic Clouds by computing pressure terms from X-ray emission, which traces shocked wind gas, and tracers of ionized gas such as radio free-free emission or optical forbidden lines \citep{lopez2011,lopez2014,mcleod2019,mcleod2020}. This is not always the case, however -- for example, \citet{pellegrini2011} concluded that winds were the dominant mechanism in the same region studied by \citet{lopez2011}. The situation is made more complex by X-ray results typically having large uncertainties compared to optical or radio measurements; furthermore, the relative scarcity of X-ray observations, combined with extinction, makes this analysis difficult even within the Milky Way \citep{barnes2020,olivier2021}. Therefore, it is still not fully certain how stellar winds and photoionization compare in terms of setting the dynamics in star-forming regions.

In a series of papers beginning with \citet{bending2020}, we investigate the intermediate scale between cloud and galaxy. We extract a section of a spiral arm from a Milky Way-like galaxy simulation, increase the resolution, and add feedback physics matching the complexity of cloud-scale models \citep[e.g.][]{dale2014}.  \citet{bending2020} detail the method for extraction and increasing resolution, and describe a ray-tracing method for photoionizing radiation emitted by cluster-sink particles. In this paper, we implement a method for stellar winds driven by ram pressure and apply it in the extracted spiral section. We compare winds with and without photoionization, providing a more detailed picture of pre-SN feedback in interacting GMCs.

%
%
\section{Numerical methods}
\label{sec:numericalmethods}

We use the smoothed particle hydrodynamics code \textsc{sphNG}, which originated with \citet{benz1990} and \citet{benz1990a}, and was substantially modified by \citet{bate1995} and \citet{price2007}. Full details of the initial conditions and cluster-sink particle setup can be found in \citet{bending2020} -- we provide a summary here. 

\subsection{Initial conditions}
\label{sec:initialconditions}
The initial conditions were extracted from a simulation by \citet{dobbs2013a} of a spiral galaxy. The galaxy was modelled using a \SI{2.5e9}{\msol} gas disc subject to a potential representing a galaxy with a two-arm spiral potential \citep{binney2008,cox2002}. This evolved for about \SI{300}{Myr} with a mass resolution of \SI{312.5}{\msol} per particle, and included self-gravity, ISM heating/cooling, H$_2$ and CO chemistry, and injections of energy representing supernova events \citep{dobbs2011}. \citet{bending2020} extracted a section of a spiral arm with dimensions $\sim 500 \times 500 \times \SI{100}{pc}$ and mass \SI{4e6}{\msol} (named `SR' in that paper). They also enhanced the resolution to \SI{1}{\msol} per particle, permitting the creation of cluster-sink particles for tracking star formation. The zoomed-in model was then evolved including self-gravity, and the same heating/cooling and chemistry as the global galaxy model \citep{glover2007,dobbs2008}. However, instead of SNe, the feedback for the zoomed-in model was photoionization from cluster-sinks. We use the same setup in this paper.

\subsection{Cluster-sink particles}

\begin{table}
	\centering
	\caption{Cluster-sink bins of stellar mass $M$, ionizing photon production rate $Q$, and wind mass loss rate \mdot{}.}
	\label{tab:sinkbins}
	\begin{tabular}{ccc} 
		\hline
		$M$ (\si{\msol}) & $\log{Q (\si{\per\s})}$ & $\log{\dot{M} (\si{\msol\per\yr})}$ \\
		\hline
19.3	&	47.7	&	-7.60	\\
21.2	&	48.0	&	-7.37	\\
23.3	&	48.3	&	-7.16	\\
25.5	&	48.5	&	-6.96	\\
28.0	&	48.6	&	-6.77	\\
30.9	&	48.7	&	-6.58	\\
34.2	&	48.9	&	-6.40	\\
37.6	&	49.1	&	-6.23	\\
41.1	&	49.2	&	-6.09	\\
45.3	&	49.3	&	-5.95	\\
50.6	&	49.4	&	-5.79	\\
56.5	&	49.5	&	-5.65	\\
62.4	&	49.6	&	-5.52	\\
69.0	&	49.7	&	-5.40	\\
87.6	&	49.9	&	-5.14	\\	
		\hline
	\end{tabular}
\end{table} 

These zoom-in models include sink particles which represent (sub-)clusters of stars. Sinks are formed according to the criteria laid out by \citet{bate1995}. For our chosen resolution, the highest density for which the Jeans mass can be resolved is \SI{1.2e4}{\per\cm\cubed}, which we set as a first density criterion for sink formation. We include a second threshold of \SI{1.2e6}{\per\cm\cubed} above which sink formation is forced. The sink accretion radius is \SI{0.78}{pc}. 

When the total mass accreted over all sinks reaches \SI{305}{\msol}, a massive star is added to whichever sink has the highest mass comprised of non-massive stars. 50 per cent of the sink mass is available for star formation. If no sink has enough mass to accept the star, the process is delayed. The massive star is taken from a pre-sampled \citet{kroupa2001} initial mass function (IMF). The ordering of stars used in this paper is the same as the models of \citet{bending2020}. See also \citealt{sormani2017} and \citealt{geen2018} for similar cluster-sink implementations. The massive stars are binned by spectral type and representative stellar properties (e.g. mass, ionizing flux) are calculated for each bin -- these are shown in \cref{tab:sinkbins}. For this paper, we create representative mass-loss rates (\mdot{}) for the stellar winds. The mass-loss rates are calculated using the MESA Isochrones \& Stellar Tracks \citep[MIST;][]{choi2016} for solar metallicity with no rotation. For each bin mass, we use the MIST tables to create an interpolated \mdot{} track over 3 Myr, and then take the mean over time as the \mdot{} for that bin. We set the wind terminal velocity $v_\infty=\SI{2000}{\km\per\s}$ in all bins.

\subsection{Photoionization}
We use a similar method as \citet{dale2007b}. Full details are provided by \citet{bending2020}. Photoionization equilibrium is calculated along lines of sight (LOS) between gas particles and ionizing sources -- the rate of photoionization due to the flux received by each particle (diluted by particles along the LOS) is balanced by the recombination rate at the density of that particle. Along each LOS, we count all particles whose smoothing length overlaps the LOS. For each particle, we treat multiple sources by adding up their individual contributions to the change in ionization fraction. Ionized gas which stops receiving ionizing radiation becomes neutral at the recombination time-scale for that density. We use the on-the-spot approximation with the case B recombination coefficient $\alpha_\textrm{B}=\SI{2.7e-13}{\cm\cubed\per\s}$, and take the ionized gas temperature to be \SI{e4}{K}. We limit the LOS to \SI{100}{pc} to alleviate the computational expense so that we can evolve models containing hundreds of ionizing sources over several Myr.

\begin{figure}
    \centering
	\includegraphics[width=0.95\columnwidth]{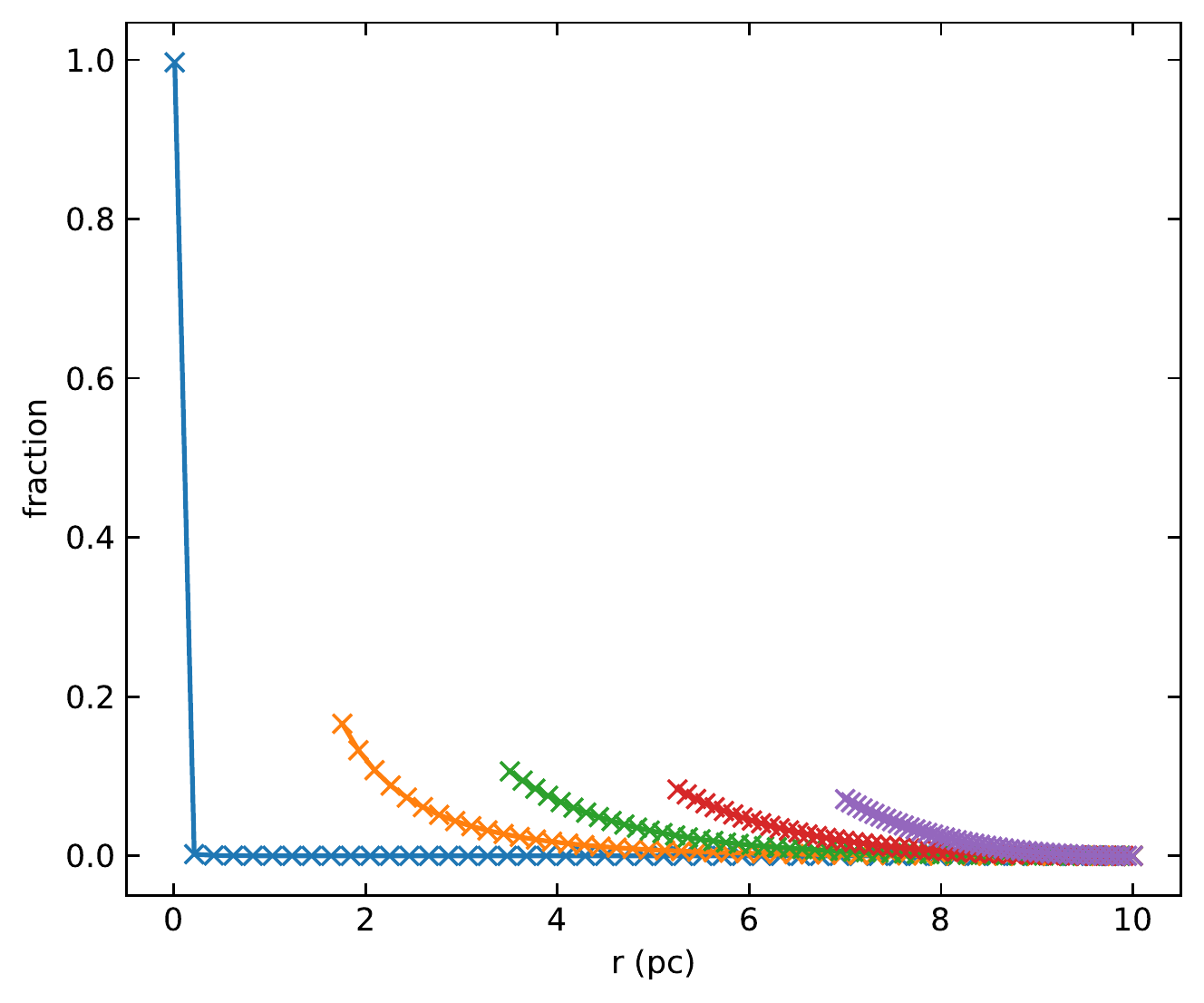}
    \caption{Toy models showing the fraction of the momentum ($w_i / \sum_j w_j$ in \cref{eq:acc}) received by 50 evenly-spaced particles with $R=\SI{10}{\pc}$ and different starting points.}
    \label{fig:weighting}
\end{figure}

\subsection{Stellar winds}

We model stellar winds as a ram pressure exerted on to gas particles near sinks. This assumes the wind bubble has cooled such that the expansion is in the momentum-conserving `snowplough' phase at the scales we simulate here. This assumption has often been taken by studies of individual clouds \citep[e.g.][]{dale2008,dale2013,dale2014,ngoumou2015,rey-raposo2017,zier2021}. Recent models by \citet{lancaster2021,lancaster2021a} show efficient radiative cooling via turbulent mixing of hot wind material with molecular clouds on the parsec scale, lending support to this assumption, especially at larger scales. 

Our implementation is based on similar methods by \citet{ngoumou2015} and \citet{rey-raposo2017}. We split the volume around each sink using the HEALPix scheme \citep{gorski2005}, which effectively creates rays emanating radially outwards from the sink. We set the number of rays, $N_\textrm{rays}$, to 48. In each ray, we identify the 50 nearest gas particles (or if there are less than 50, for example if the sink is near the edge of the computational volume, then we select all particles). The number of particles selected in a ray is $N_\textrm{w}$. The total wind momentum per unit time ($\dot{M} v_\infty$) is distributed evenly over the rays, and then distributed over the selected particles in each ray according to a weighting factor which depends on distance to the sink, with nearer particles receiving a larger fraction of the ray momentum. The weighting factor for particle $i$ is
\begin{equation}
    \label{eq:weight}
    w_i = \frac{1}{r_i^2}\frac{(r_i-R)^2}{R^2}
\end{equation}
where $r_i$ is the distance to particle $i$, and $R=r_{N_\textrm{w}}$ (the distance to the furthest selected particle in the ray).
The force exerted on particle $i$ is then
\begin{equation}
    \label{eq:acc}
    m_i a_i = \frac{\dot{M} v_\infty}{N_\textrm{rays}} \frac{w_i}{\sum_{j=1}^{N_\textrm{w}} w_j} .
\end{equation}
The effect of the normalised weighting factor is shown in \cref{fig:weighting} for a toy model with 50 evenly spaced particles. Particles closer to the origin receive a larger share of the momentum; the closer a particle is to the origin, the more pronounced this becomes. This results in a wind bubble which expands from the inside out. As with the photoionization algorithm, we only apply winds within a distance limit of \SI{100}{pc} around a sink.

\subsubsection{Single star in a uniform medium}

\begin{figure}
    \centering
	\includegraphics[width=0.95\columnwidth]{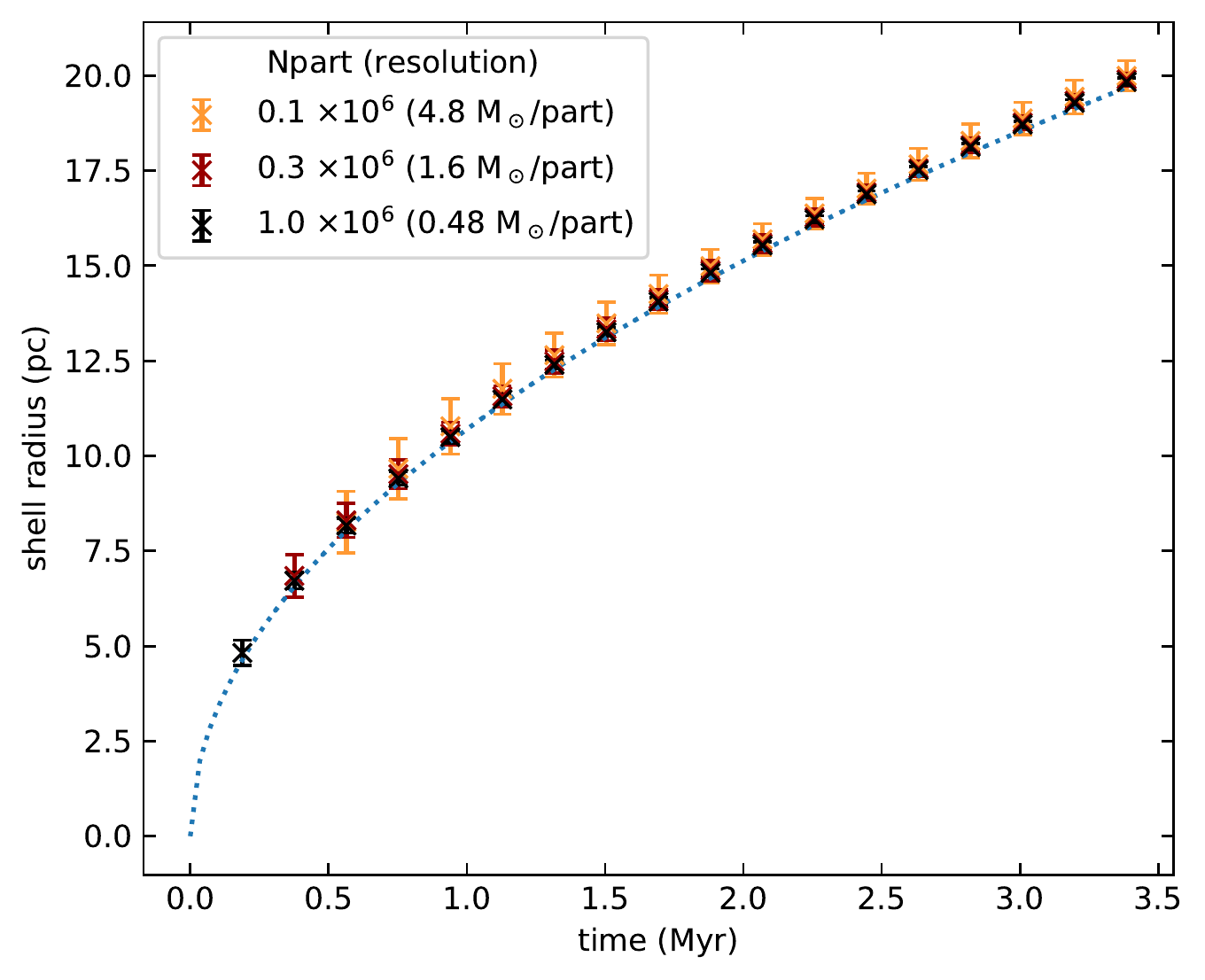}
    \caption{Expansion of a thin shell in a uniform-density medium with constant wind mass-loss rate. We test three particle resolutions. The line shows the analytical solution.}
    \label{fig:resolution}
\end{figure}

To test the stellar wind method, we place a single star in the centre of a uniform-density cloud. The wind ram pressure drives the expansion of a thin shell through the cloud -- from momentum conservation, the shell position as a function of time is given by 
\begin{equation}
    \label{eq:rshell}
    R(t) = \left( \frac{3}{2\pi}\frac{\dot{M} v_\infty}{\rho_0} \right)^{1/4} t^{1/2}
\end{equation}
\citep{lamers1999,capriotti2001} where $\rho_0$ is the initial density of the ambient medium, \mdot{} is the wind mass-loss rate, and $v_\infty$ is the wind terminal velocity. We place the star in a cloud of mass \SI{5e5}{\msol}, radius \SI{54.42}{pc}, density $\rho_0=\SI{30}{\per\cm\cubed}$, and temperature \SI{10}{K} (with an isothermal equation of state). The wind parameters are $\dot{M} = \SI{e-5}{\msol\per\yr}$ and $v_\infty=\SI{2000}{\km\per\s}$. 
The result of the test at different resolutions is shown in \cref{fig:resolution}. The shell radius is calculated by taking a density-weighted mean of particle positions which have $\rho > 1.2 \rho_0$. The error bar distance in each direction is half the distance between the shell radius and cavity radius. Points are absent if the shell radius is not well defined (i.e. there are no particles with density above $1.2 \rho_0$).
All particle resolutions track the time-evolution accurately. However, it takes more time at lower resolutions for the thin dense shell to become well defined by our definition. 

\subsection{Spiral section model parameters}

We compare stellar winds with photoionization in the spiral arm section described in \cref{sec:initialconditions}. One model contains both processes, and one model includes just winds. We compare our results with the equivalent models by \citet{bending2020}: one with just photoionization (named in that paper as `SR\_50\%'), and one with no feedback (`SR'). The results are presented in the following section.

%
%
\begin{figure*}
    \centering
	\includegraphics[width=0.95\textwidth]{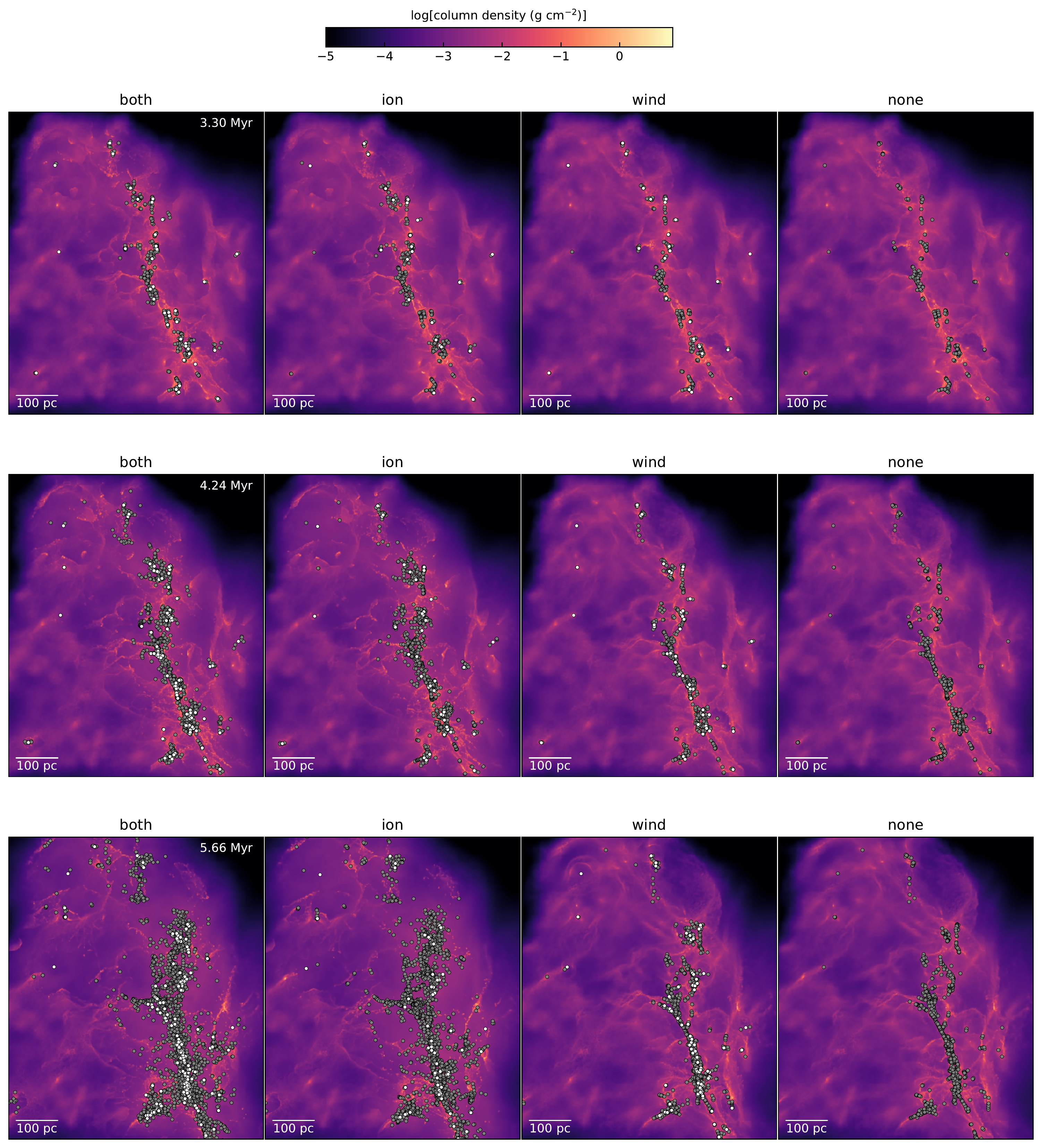}
    \caption{Top-down view ($x$-$y$ plane) of column density in the four models at three different times. Dots are sink particles; white dots produce stellar feedback. }
    \label{fig:colden}
\end{figure*}

\begin{figure*}
    \centering
	\includegraphics[width=0.95\textwidth]{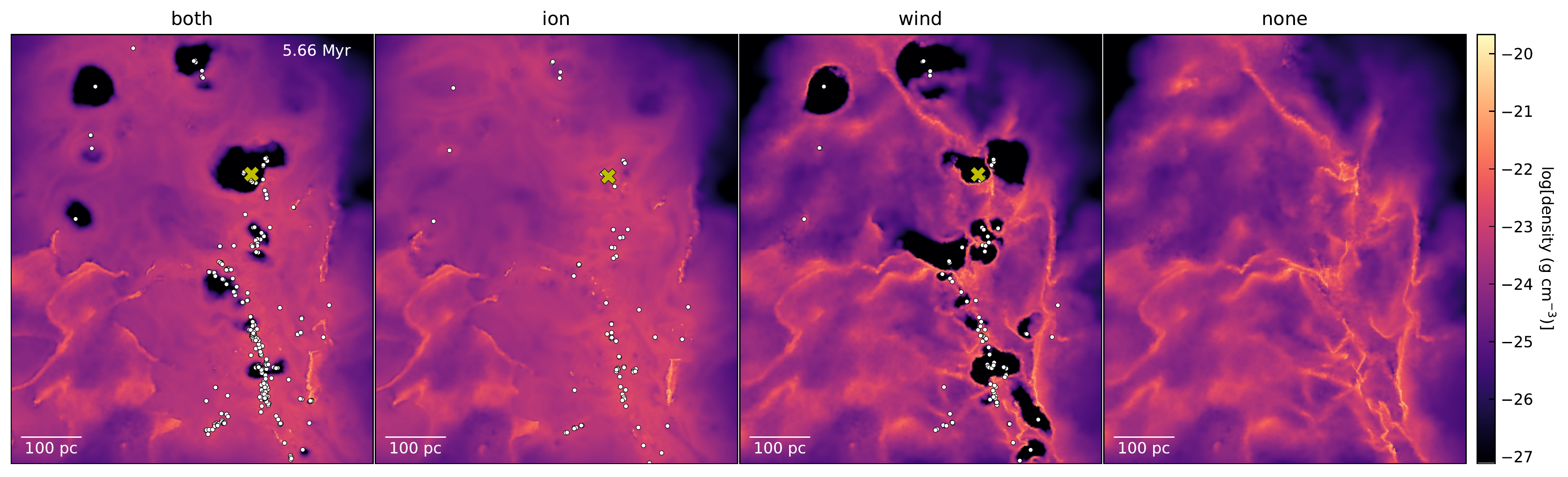}
    \caption{Density cross-section at $z=0$, for the third row of \cref{fig:colden}. Only the sinks producing feedback are shown. The yellow X shows the sink referred to in \cref{sec:impact}/\cref{fig:rimpact_time}.}
    \label{fig:xsec}
\end{figure*}

\begin{figure*}
    \centering
	\includegraphics[width=0.95\textwidth]{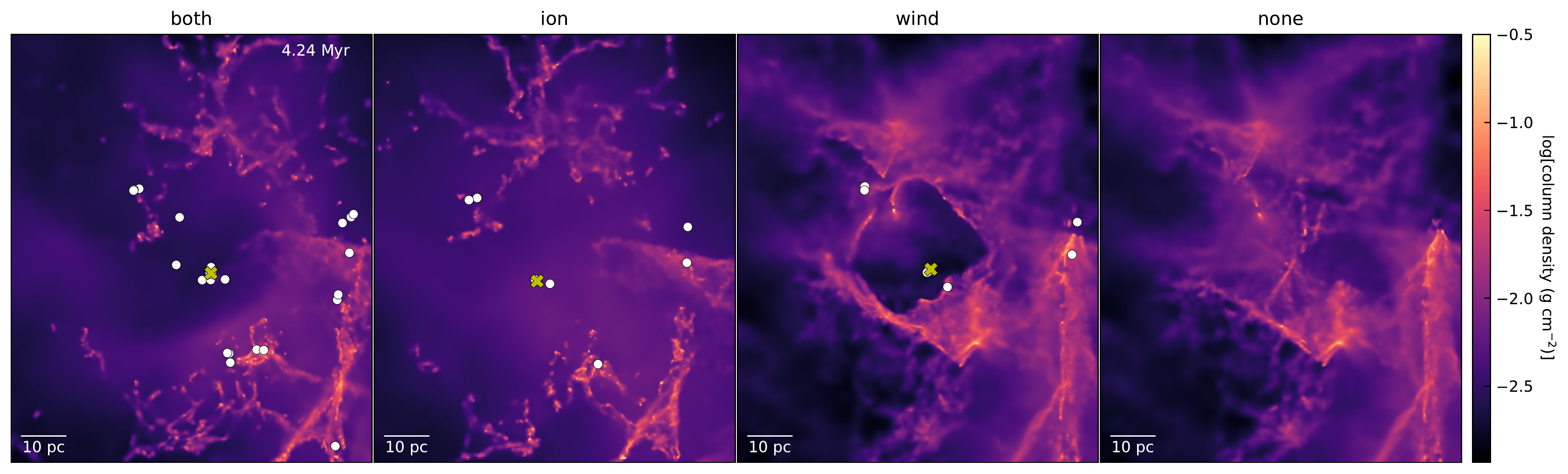}
    \caption{Column density in a star-forming region. Ionization disrupts the whole region, while stellar winds create small-scale bubbles. White dots show sinks producing feedback. The yellow X shows the sink referred to in \cref{sec:impact}/\cref{fig:rimpact_time}.}
    \label{fig:region2}
\end{figure*}

\section{Results}
\cref{fig:colden,fig:xsec} show snapshots of column density and density cross-section, respectively, at three different times. Ionization breaks up material in the spiral arm, and makes the diffuse gas smoother. Models with ionization have sink particles more spread out compared to models without ionization. The model with just winds is very similar to the no-feedback model, except with bubbles around massive stars (particularly seen in the cross-section images). The effect is similar for the model with both feedback mechanisms -- winds create bubbles whilst ionization is responsible for disrupting the spiral arm. Thus stellar winds have only a small-scale effect on the morphology of the gas, whereas the impact of ionization is felt over the whole region. This also means wind bubbles require cross-sections to be easily identified -- they are not readily apparent in column density, except when the bubble is relatively isolated (such as the arcs in the top left of the `wind' column of \cref{fig:colden}). When ionization heating is not included, the wind shell is denser and thinner. 

The differences in morphology can be seen more clearly in \cref{fig:region2}, which focuses on a particular star-forming region. The wind-only model forms a well-defined shell around a cleared-out bubble surrounding the central cluster. The ionization-only model, however, is able to disrupt the entire region, including the neighbouring filamentary structure to the right of the frame. High-density gas is also disrupted, with the morphology taking the form of knots rather than coherent filaments. This model contains smooth, diffuse gas around the cluster, while the wind model has a more excavated bubble bounded by sharp density gradient. The model with both feedback mechanisms most closely resembles the ionization-only model, with the addition of a cavity in the diffuse, ionized gas. In this case, the wind cavity is not bounded by a sharp density gradient, as the external medium has a larger pressure -- i.e. the wind ram pressure is going into a warm, ionized medium instead of a cold, neutral one. When put together, photoionization dominates the evolution of the region, while stellar winds only affect the ionized gas component. This behaviour is also seen by \citet{dale2014}, who modelled the two feedback processes in individual molecular clouds. 

\begin{figure}
    \centering
	\includegraphics[width=0.95\columnwidth]{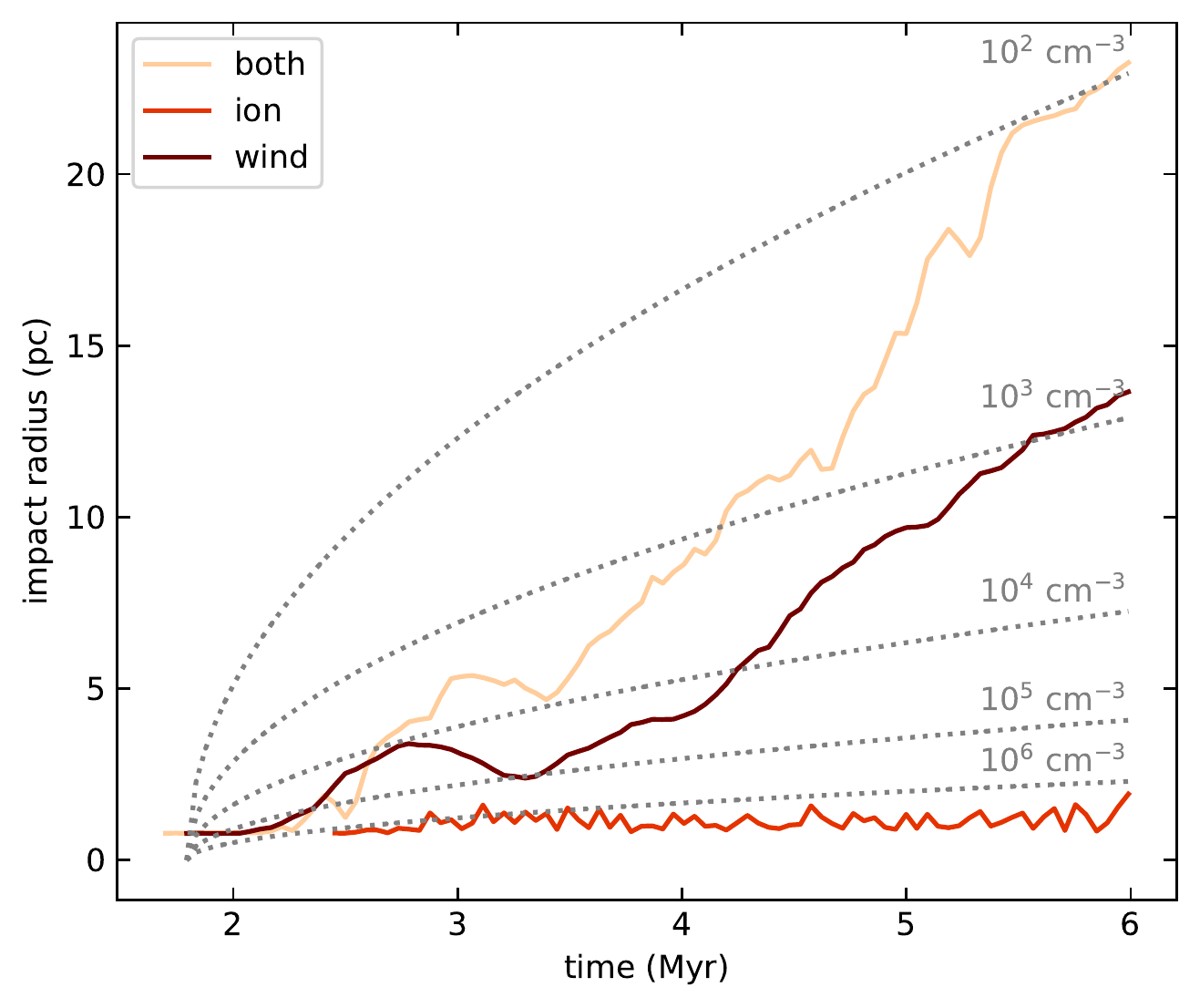}
    \caption{Time-evolution of the wind impact radius, $R_\textrm{imp}$, for the sink particle with (approximately) the highest $\dot{M}$ at \SI{5.66}{\mega\yr}  (see \cref{sec:impact}). The ionization-only model is also shown for comparison -- in this case, the sink is the one with the highest ionizing flux. Lines start at the time massive stars form for that sink. The dotted lines show \cref{eq:rshell} for different $\rho_0/m_\textrm{H}$, using a constant $\dot{M} = \SI{4e-5}{\msol\per\yr}$. They are shown as points of reference only -- the actual regions have inhomogeneous density profiles and experience increasing \mdot{}.}
    \label{fig:rimpact_time}
\end{figure}

\begin{figure}
    \centering
	\includegraphics[width=0.95\columnwidth]{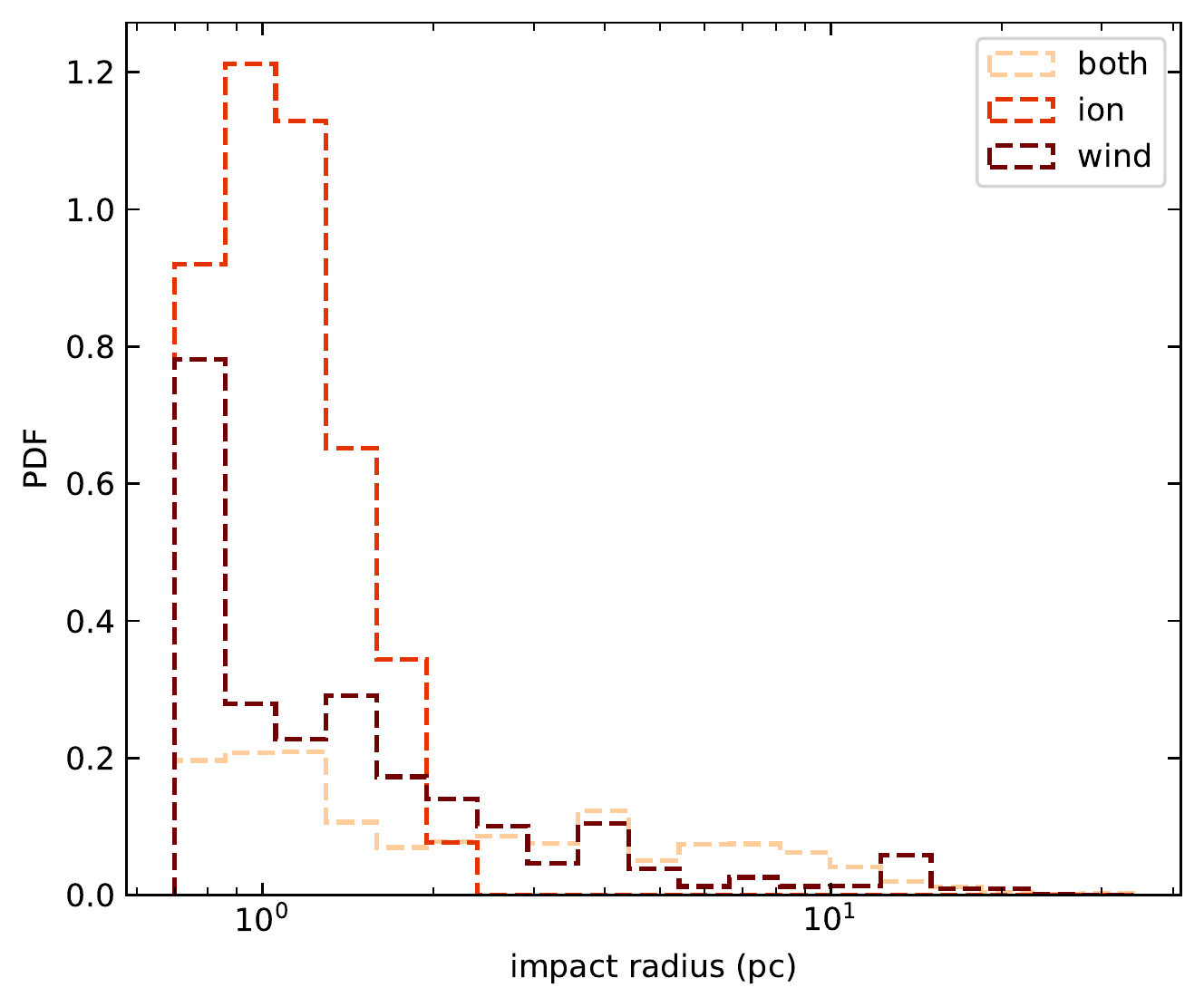}
    \caption{PDF of $R_\textrm{imp}$ for all feedback-producing sinks at \SI{5.66}{\mega\yr} (see \cref{sec:impact}).}
    \label{fig:rimpact_pdf}
\end{figure}

\subsection{Wind bubbles}
\label{sec:impact}

Stellar winds create cavities around massive stars, which can been seen in the density maps in \cref{fig:xsec}. In this section, we characterise the size and evolution of these cavities. 

At \SI{5.66}{\mega\yr}, we locate the sink particle which has the largest wind $\dot{M}$ in the combined-feedback model (\SI{5.8e-5}{\msol\per\yr}). We then find the equivalent sink in the other feedback models such that the same region of space is analysed -- in the ionization-only model, this is the sink with the highest flux (\SI{1.5e51}{\per\s}), and in the wind-only model, it is the sink with the second largest $\dot{M}$ (\SI{5.4e-5}{\msol\per\yr}). The sink positions are marked with a yellow X in \cref{fig:xsec,fig:region2}. We track these sinks at earlier and later times, and estimate the cavity sizes.

We define the impact radius, $R_\textrm{imp}$, as the distance between the tracked sink and its nearest gas particle. This is the smallest radius where the stellar wind first collides with the ISM (or would do if winds were switched on). The time-evolution of $R_\textrm{imp}$ is shown in \cref{fig:rimpact_time}. The dotted lines show \cref{eq:rshell}, which assumes the expansion of a spherically symmetric bubble into a uniform density medium with a constant $\dot{M}$. They are provided as idealised points of reference to compare with the measured $R_\textrm{imp}$ -- the actual regions have inhomogenous density profiles and varying \mdot{}, so the measured results are not expected to follow these analytical lines. Lines are shown for densities $\rho_0/m_\textrm{H}$ between $10^2$ and \SI{e6}{\per\cm\cubed}, with $\dot{M} = \SI{4e-5}{\msol\per\yr}$ (which is the mean $\dot{M}$ over time for the chosen sink particles; for simplicity in the figure, we do not plot lines of different $\dot{M}$). 

The combined-feedback model grows the largest cavity, with $R_\textrm{imp}= \SI{23}{\pc}$ after approximately \SI{4}{\mega\yr} of evolution. The wind-only model ends with a smaller cavity of size \SI{14}{\pc}. The final radii are similar to what would result from evolving a constant wind in densities of $10^2$ and \SI{e3}{\per\cm\cubed}, respectively. The ionization-only model has a negligible cavity throughout ($< \SI{2}{\pc}$).

\cref{fig:rimpact_pdf} shows the probability density function (PDF) of $R_\textrm{imp}$ for all feedback-producing sink particles at \SI{5.66}{\mega\yr}. The PDF is normalised by the total number of feedback sinks multiplied by the bin width, with the integral of the PDF equalling unity.  Note that this is not a distribution of cavity sizes, but of the distance to the first deposition of wind momentum for all feedback sinks (some sinks may lie in the same cavity). As with \cref{fig:rimpact_time}, the results for ionization-only model show where winds would be deposited if they were switched on. $R_\textrm{imp}$ remains below \SI{2}{pc} for all sinks in the ionization-only model, while models with winds produce much larger $R_\textrm{imp}$, going up to \SI{32}{\pc}. The wind-only model is skewed towards smaller $R_\textrm{imp}$ (median \SI{3.1}{\pc}) compared to the combined-feedback model (median \SI{5.9}{\pc}). The minimum $R_\textrm{imp}$ for each model corresponds to the sink accretion radius (\SI{0.78}{\pc}). 

These results show that stellar winds are able to clear gas away from the vicinity of massive stars -- up to tens of pc at the most extreme -- while ionization by itself is not.

\subsection{Star formation}

\begin{figure*}
    \centering
	\includegraphics[width=0.70\textwidth]{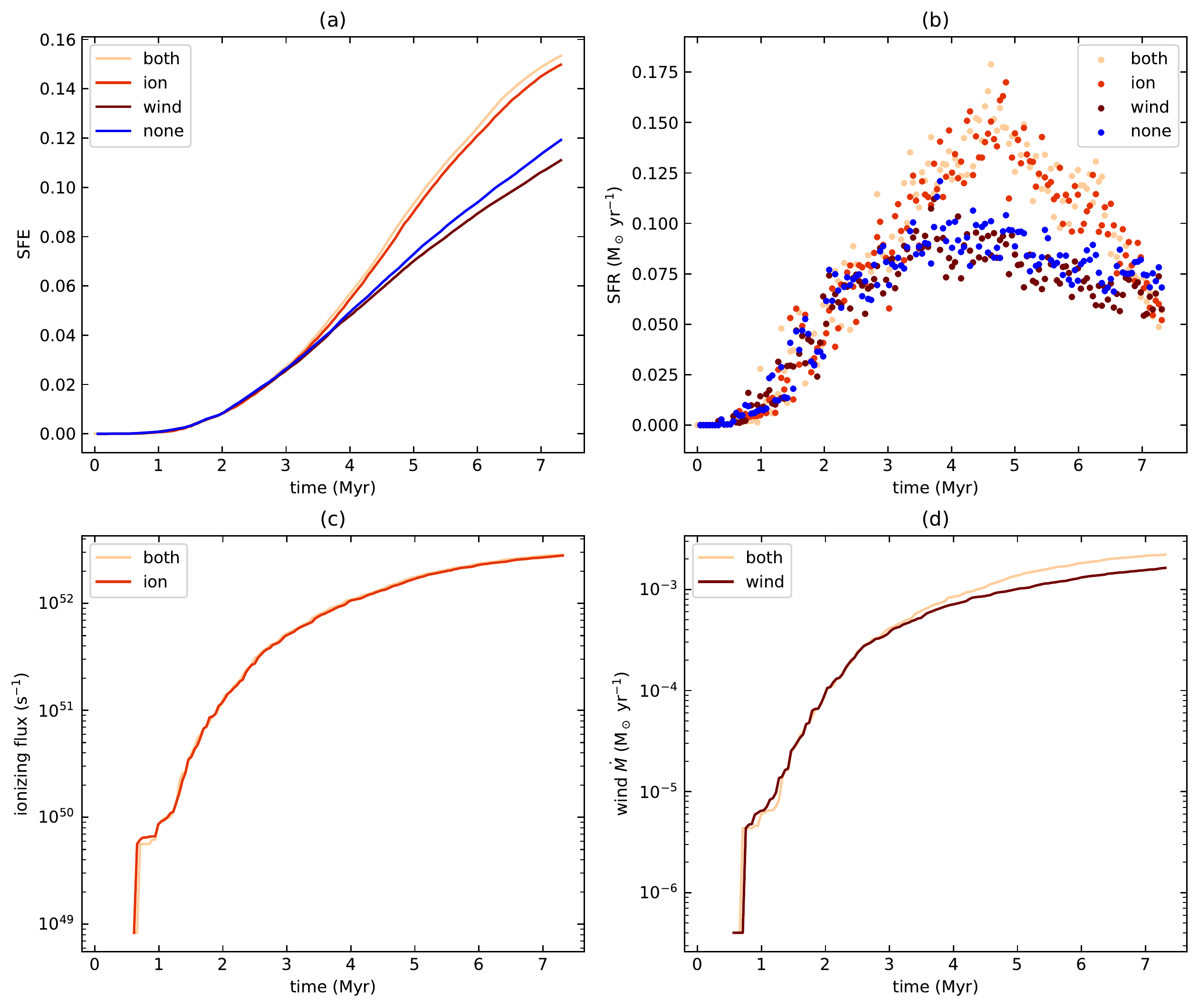}
    \caption{(a) Star formation efficiency, (b) star formation rate, (c) integrated ionizing flux, (d) integrated wind mass loss rate.}
    \label{fig:sfe}
\end{figure*}
\begin{figure}
    \centering
	\includegraphics[width=0.95\columnwidth]{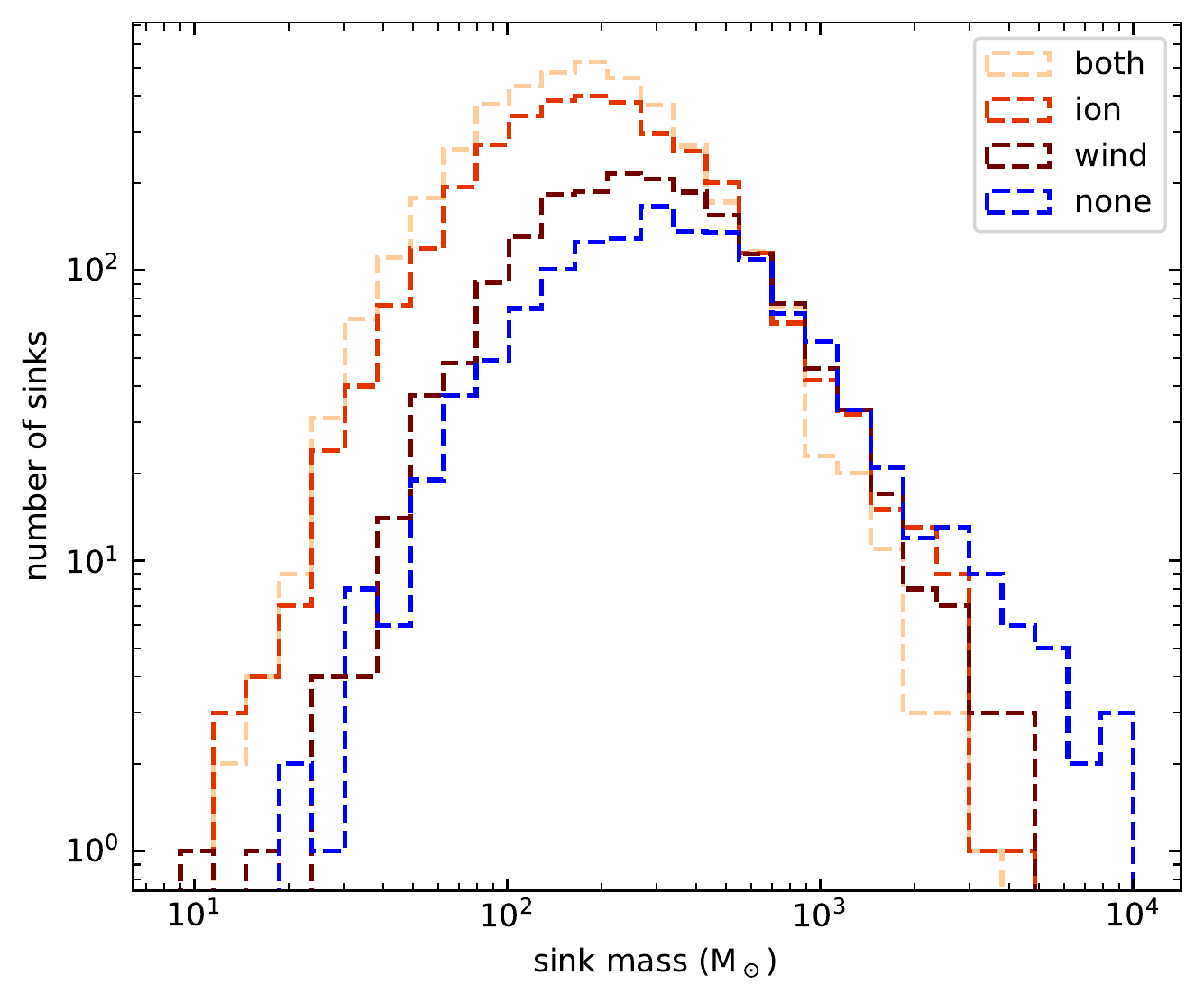}
    \caption{Histogram of sink masses at \SI{5.66}{\mega\yr}.}
    \label{fig:sinkmass}
\end{figure}

\cref{fig:sfe} shows properties of sink particles as a function of time. Panel (a) shows the total star formation efficiency (SFE), defined as
\begin{equation}
    \label{eq:sfe}
    \mathrm{SFE} = 0.5 \times \frac{M_\mathrm{sinks}}{M_\mathrm{sinks} + M_\mathrm{gas}} .
\end{equation}
The factor of 0.5 is the proportion of the sink mass which is converted into stars. After 7.3 Myr, the highest SFE is 0.15 which occurs for the combined and ionization-only models. The model with no feedback has SFE = 0.12, while the wind-only model has the lowest SFE of 0.11. Therefore, the models which include ionization moderately boost star formation. The impact of stellar winds is marginal, only changing the SFE by 0.01 at most.

Panel (b) of \cref{fig:sfe} shows the star formation rate (SFR), defined as
\begin{equation}
    \label{eq:sfr}
    \mathrm{SFR}(t_n) = 0.5 \times \frac{M_\textrm{sinks}(t_n) - M_\textrm{sinks}(t_{n-1})}{t_{n} - t_{n-1}}
\end{equation}
where $t_{n-1}$ and $t_{n}$ are two consecutive dump times. The four models progress at approximately the same rate, diverging at approximately \SI{3.7}{Myr} when the wind-only and no-feedback models reach their peak SFR. The ionization-only and combined-feedback models continue to increase, peaking at approximately \SI{4.7}{Myr}. By the end of the runtime, at \SI{7.3}{Myr}, all four models have converged to the same SFR again. As with the SFE, stellar winds have a negligible affect on the SFR, while ionization allows the SFR to reach a higher peak than the other models (0.15 vs \SI{0.10}{\msol\per\yr}, respectively). 

Panel (c) of \cref{fig:sfe} shows the total ionizing flux integrated over all sink particles. The total flux is indistinguishable between the two models which include ionization. Similarly, the total wind mass-loss rates shown in panel (d) are effectively the same for the two wind models. 

\cref{fig:sinkmass} shows a histogram of sink masses at \SI{5.66}{\mega\yr}. The control run without feedback produces more high-mass sink particles ($>$\SI{e3}{\msol}) compared to the models with either feedback mechanism -- ionization and winds prevent the most massive sinks forming, particularly when combined together. Instead, the feedback runs produce more low-mass sinks below $\sim$\SI{500}{\msol}. Ionization has the most drastic impact, with stellar winds appearing as a second-order effect.

It is clear from the SFE and SFR that photoionization drives additional star formation on a global scale in our models \citep[see also][]{bending2020}, while stellar winds have a negligible effect on these integrated quantities. However, winds do affect how the star formation is distributed, spreading the mass over more low-mass sink particles and producing fewer of the heaviest sinks, due to the formation of wind-blown bubbles (see \cref{sec:impact}).

\subsection{Gas kinematics}

\begin{figure}
    \centering
	\includegraphics[width=0.95\columnwidth]{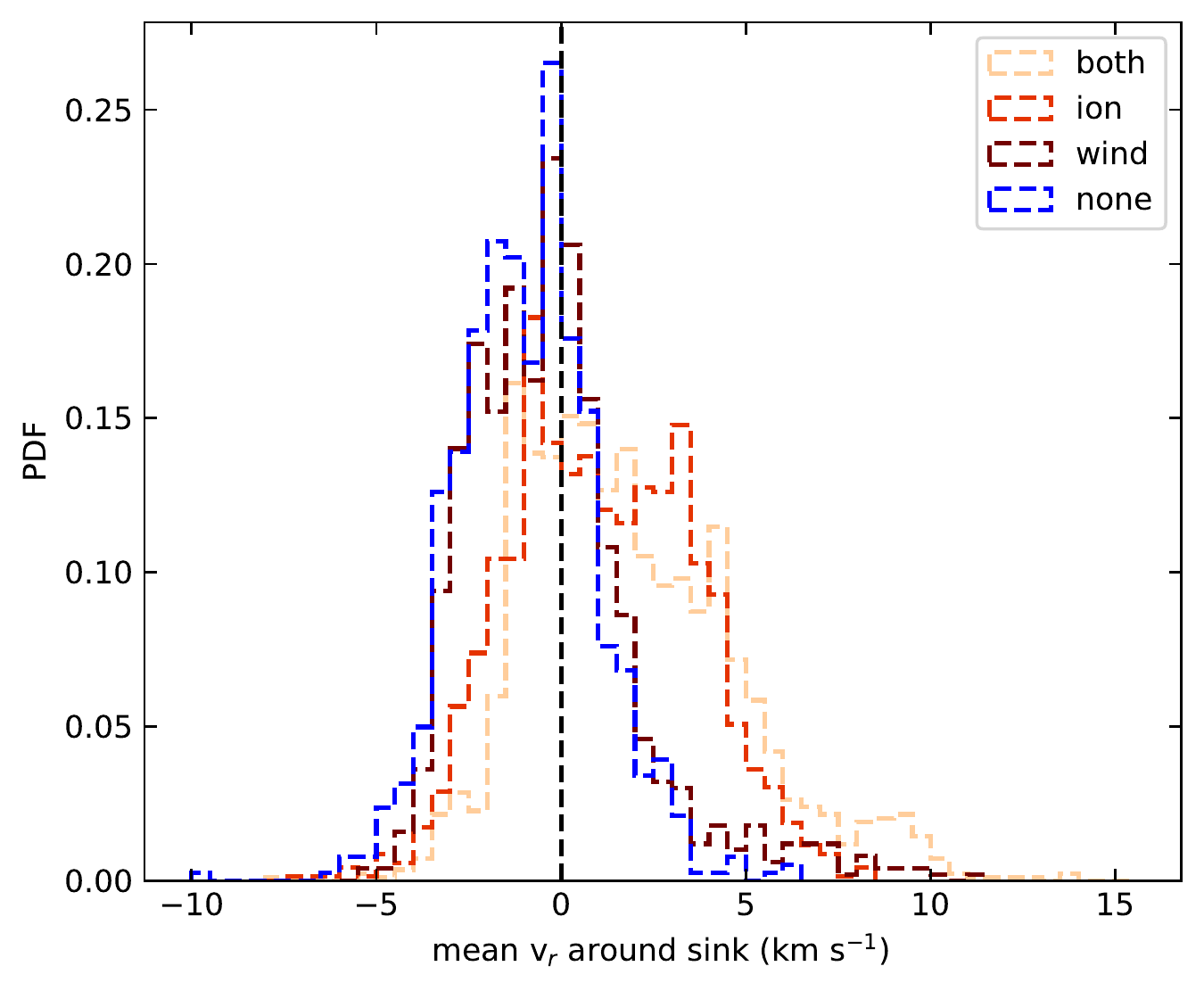}
    \caption{PDF of the mean radial velocity within \SI{20}{pc} around sink particles at \SI{4.24}{\mega\yr}.}
    \label{fig:vrad}
\end{figure}

We investigate the ability of each feedback mechanism to expel gas near sink particles. For all particles $i$ inside a radius of \SI{20}{pc} around a sink, we calculate the particle's radial velocity $v_{r,i}$,
\begin{equation}   
   \begin{split}    
   v_{r,i} &= (\mathbfit{v}_i - \mathbfit{v}_*) . \hat{\mathbfit{r}} \\
     &= (\mathbfit{v}_i - \mathbfit{v}_*) .  \left(\frac{\mathbfit{r}_i - \mathbfit{r}_*}{|\mathbfit{r}_i - \mathbfit{r}_*|} \right)
   \end{split}
\end{equation}
i.e. $v_{r,i}$ is the component of the particle velocity, relative to the sink velocity, in the direction pointing radially away from the sink. We take the mean of all the $v_{r,i}$ around that sink. This is repeated for all sinks. We then plot a PDF of all the means, shown in \cref{fig:vrad} at time \SI{4.24}{\mega\yr}. The wind-only and no-feedback models have peak mean radial velocities below \SI{0}{\kms} (i.e. implying most of the sinks have infall in their vicinity). On the other hand, the majority are positive in the ionization-only and combined-feedback models -- that is, including ionization shifts the PDF to the right towards higher velocities. These results provide a global, averaged picture of gas kinematics and imply that ionization is able to expel gas, while still allowing significant infall to occur; stellar winds, however, are negligible according to this measure. The velocities also help explain the distribution of sink masses in \cref{fig:sinkmass} -- higher mean radial velocities around sinks in the ionization models results in lower mass sinks compared to the models without ionization.

\begin{figure*}
    \centering
	\includegraphics[width=0.95\textwidth]{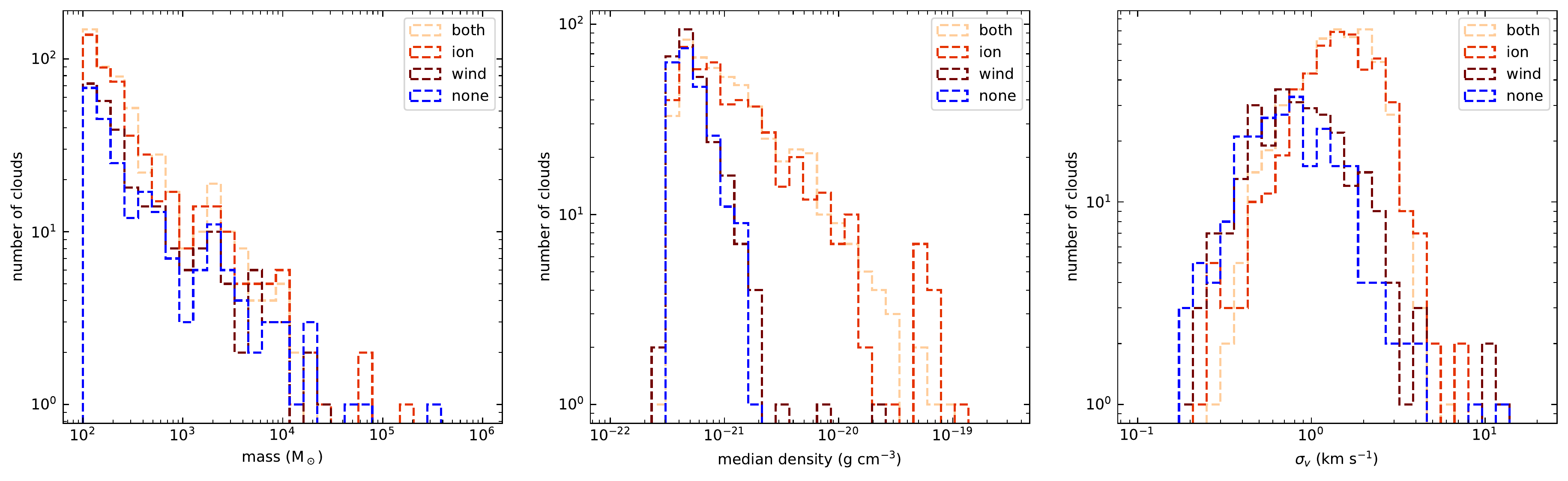}
    \caption{Histograms of cloud mass, median density, and velocity dispersion at \SI{4.24}{Myr}.}
    \label{fig:cloudhist}
\end{figure*}

\subsection{Clouds}

We identify clouds using a friends-of-friends algorithm, which groups particles together if their nearest neighbours are within a specified distance of each other. We require a cloud to have a minimum of 100 particles and a maximum particle separation of \SI{0.55}{pc}, which are the same parameters used by \citet{bending2020} (whose fig. 11 shows the cloud structures in ionization and no-feedback models). Additionally, we require all chosen particles to be neutral. 

At \SI{4.24}{\mega\yr}, the total number of clouds are 510 (both processes), 471 (ion), 271 (wind), and 232 (no feedback), indicating that stronger feedback results in more clouds. In particular, including ionization doubles the number of clouds compared to the no-feedback case, while adding winds creates a small number of additional clouds (around 10 to 15 per cent more compared to the models without winds). The impact of ionization can be seen in the density figures of \cref{fig:colden} and \cref{fig:xsec}, which show the spiral arm being broken apart and gas being collected into dense shells, which increases the number of detected clouds. 

\cref{fig:cloudhist} shows histograms of the cloud mass, median density, and the standard deviation of the velocity. 
A power-law fit is calculated for the mass histogram below \SI{e4}{\msol} using a least-squares method. The derivative is then calculated to find the mass function in the form $\dif{N}/\dif{M} \propto M^\gamma$, where the indices are $\gamma =$ 
$-1.77 \pm 0.07$ (both), 
$-1.72 \pm 0.07$ (ion), 
$-1.70 \pm 0.08$ (wind), and 
$-1.66 \pm 0.09$ (none) 
-- there is no significant difference between the shape of the mass functions. These values agree with the index of around $-1.7$ for clumps observed in the Milky Way \citep{solomon1987,heyer2001,roman-duval2010,colombo2019,ma2021}, as well as simulations of Milky Way-like galaxies \citep{dobbs2011,jeffreson2021a}, although the mass ranges may differ depending on resolution (low mass clouds are more difficult to measure).

The density histograms show larger differences between the models, with the non-ionizing models having a narrower distribution which stops at \SI{e-21}{\g\per\cm\cubed}, while the ionization models extend to \SI{e-19}{\g\per\cm\cubed}. The velocity dispersions are higher in the ionization models, with median values of \SI{1.4}{\kms}, compared to \SI{0.8}{\kms} in the models without ionization. The velocity dispersions found in clouds with ionization agree better with typical results for observed clouds of similar masses \citep{roman-duval2010,ma2021,duarte-cabral2021}.

Including winds does not produce a significant difference in any of the histograms; the largest difference between the ionization-only and combined feedback models occurs at the tail end of the density distribution around \SI{3e-20}{\g\per\cm\cubed}, but definitive conclusions can not be drawn due to the small number of clouds here. 

In summary, stellar winds have minimal impact, limited to creating around 10 per cent more clouds than models without winds. Ionization plays a greater role, producing twice as many clouds which tend towards higher densities and velocity dispersions.

\subsection{Sink clustering}
\begin{figure}
    \centering
	\includegraphics[width=0.95\columnwidth]{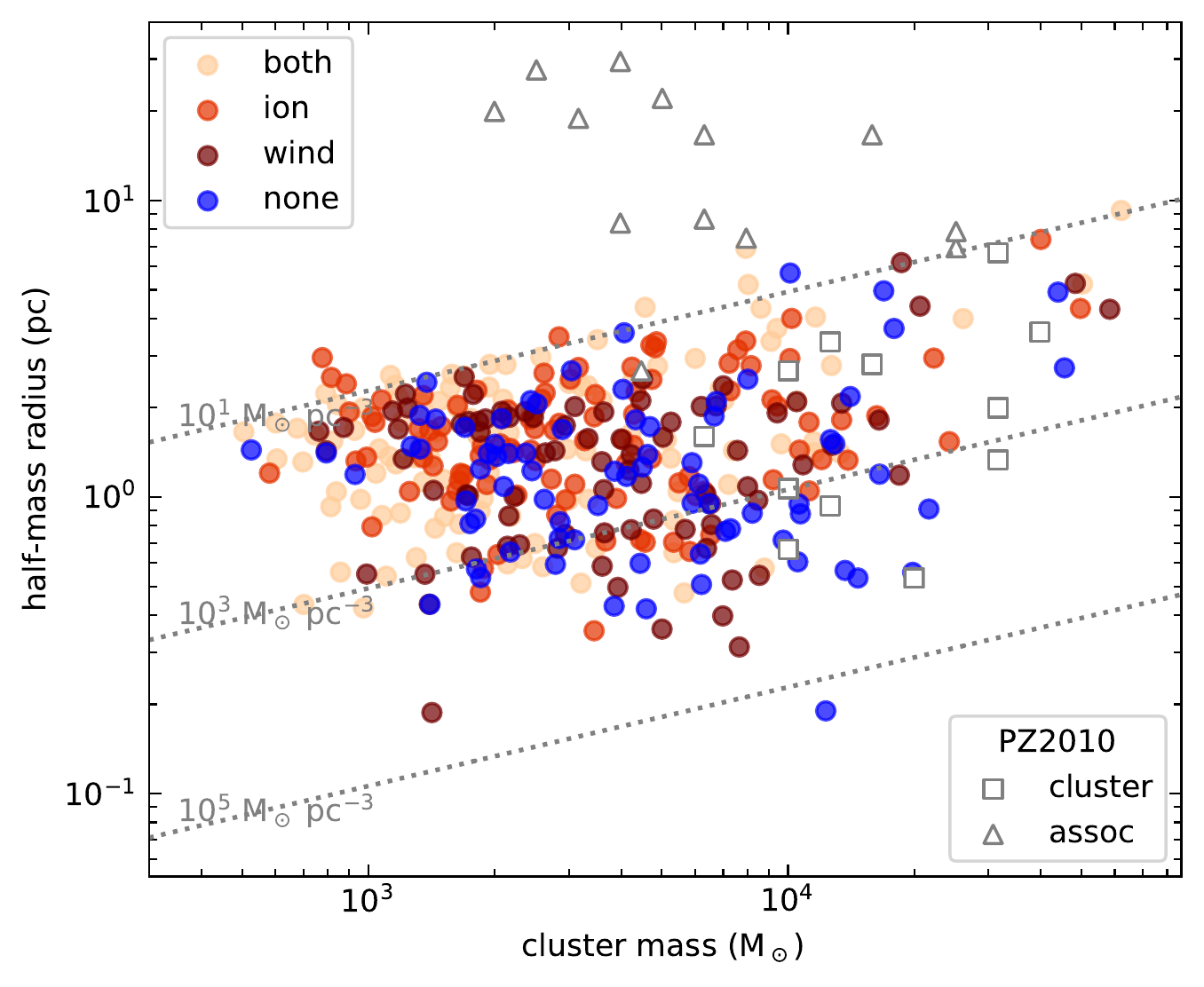}
    \caption{Properties of clusters identified with DBSCAN at \SI{5.66}{Myr}. Milky Way YMCs (squares) and associations (triangles) from \citet{portegies-zwart2010} are shown for comparison. Lines of constant half-mass density in \si{\msol\per\pc\cubed} are also shown.}
    \label{fig:rhm}
\end{figure}

\begin{figure}
    \centering
	\includegraphics[width=0.95\columnwidth]{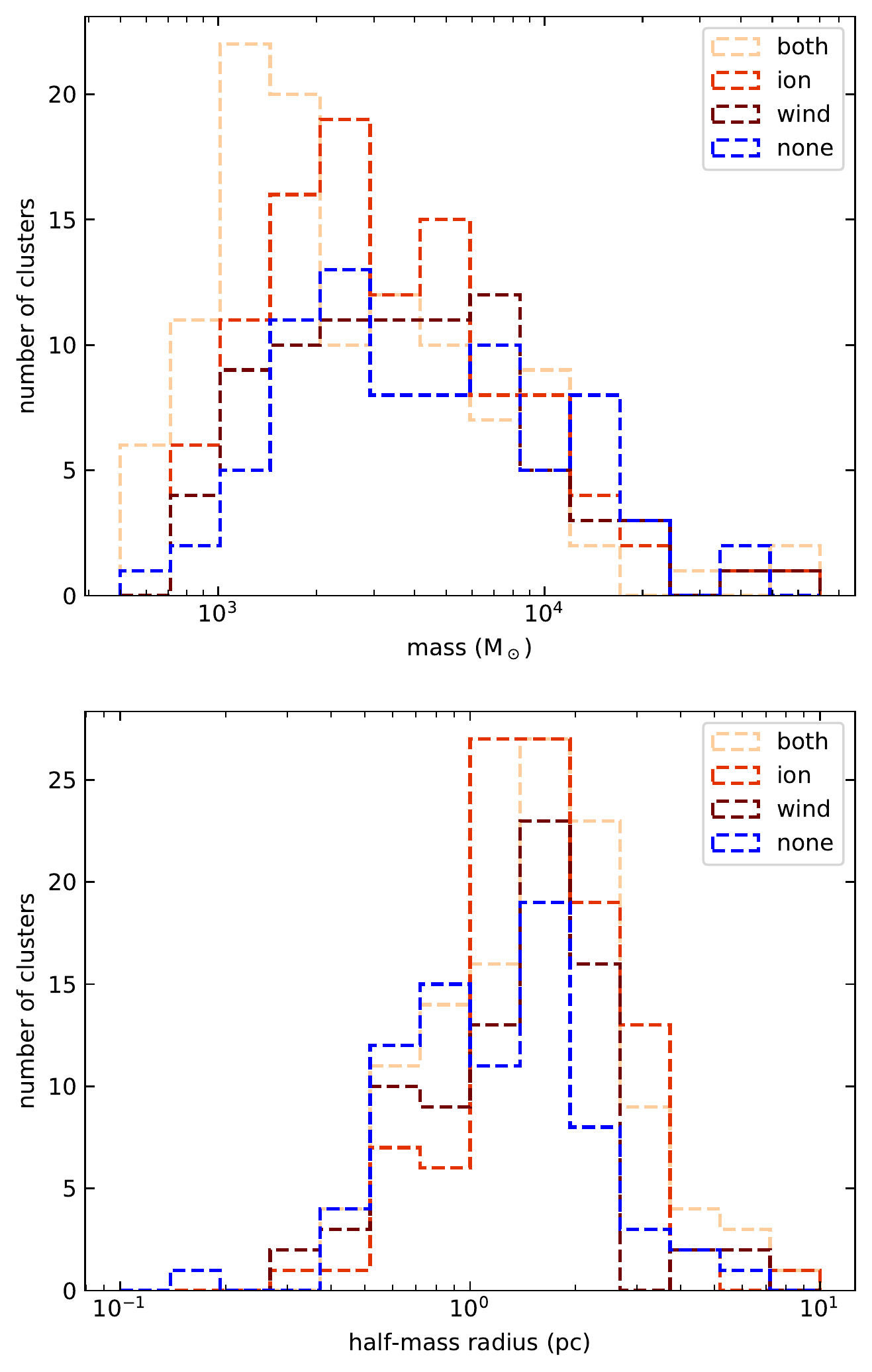}
    \caption{Histograms of mass and half-mass radius for clusters identified with DBSCAN at \SI{5.66}{Myr}.}
    \label{fig:dbscan_hist}
\end{figure}

At the resolution of these models, a sink particle represents a (sub)cluster of stars rather than individual stellar particles. Although the precise properties of stellar clusters are therefore not resolved, we attempt an analysis of the spatial clustering of these sink particles and compare them with clusters in the Milky Way which host massive stars.  

We use the DBSCAN algorithm \citep[Density-Based Spatial Clustering of Applications with Noise;][]{ester1996} -- see e.g. \citet{joncour2018} for a comprehensive description of the method and its usage for observed stellar systems, and \citet{liow2020} for an application in simulations of cloud-cloud collisions. We set 5 as the minimum number of members required to define a cluster, and $\epsilon = \SI{3}{\pc}$ as the maximum neighbour separation. This parameter is chosen by calculating the distance to the 5th nearest neighbour of each sink particle, and plotting the sorted results against the sink indices. The plot rapidly diverges at the optimal $\epsilon$ \citep{rahmah2016}. 

\cref{fig:rhm} shows the cluster mass, $M$, against half-mass radius, $r_\textrm{hm}$ (the distance away from the centre of mass which contains half the cluster mass). For comparison, we also show the Milky Way young massive clusters (YMCs) and associations from table 2 of \citet{portegies-zwart2010}. Lines of constant half-mass density ($3M/8\upi r_\textrm{hm}^3$) are also shown. The model results are denser than \SI{10}{\msol\per\pc\cubed}, consistent with the observations of YMCs as opposed to associations, most of which lie at lower densities. All the feedback models occupy the full span of the parameter space as the no-feedback case; for example, feedback (or lack thereof) does not prevent the formation of YMCs. However, both feedback mechanisms produce more low-mass clusters. 

This can be seen more clearly in \cref{fig:dbscan_hist}, which shows histograms of the cluster mass and half-mass radius. Unlike with gas clouds, stellar winds do have an effect on the cluster properties, especially when combined with ionization -- the two processes together produce more low-mass clusters than either winds or ionization individually. This is due to more low-mass sink particles being produced, as shown in \cref{fig:sinkmass}. Fewer small clusters (radii $< \SI{1}{pc}$) are created, while a greater number of large clusters ($> \SI{1}{pc}$) are produced when feedback is included. 

Observations and simulations show that cluster mass functions follow a power law \citep{lahen2020,hislop2021}, possibly with an exponential tail at high masses \citep{portegies-zwart2010,li2017}. However, for our models, the small number of clusters and the large scatter between mass bins makes the mass function more difficult to calculate for clusters than for clouds. Therefore, we leave this for future studies where we aim to improve how stars are resolved in clusters.

%
%
\section{Discussion and conclusions}
\label{sec:conclusions}
We have presented SPH simulations of photoionization and stellar winds in a $500\times500\times\SI{100}{\pc}$ section of a spiral arm, building on the work by \citet{bending2020}. The initial conditions were extracted from a galaxy simulation \citep{dobbs2013a} and the resolution enhanced to \SI{1}{\msol} per particle. The feedback implementation is of similar robustness to that included in models of individual clouds on scales of a few tens of pc \citep[e.g.][]{dale2014}, but here applied to larger scales. In summary, our key results are:
\begin{enumerate}
    \item Photoionization is the dominant feedback mechanism which disrupts the spiral arm section, while stellar winds play a negligible role.
    \item Stellar winds do not affect the SFE or SFR.
    \item However, each mechanism affects the distribution of star formation, producing more low-mass sinks and fewer high-mass sinks ($>$\SI{e3}{\msol}).
    \item The main morphological impact of stellar winds is the formation of small-scale cavities ($\sim$10--\SI{30}{\pc})
    \item Both feedback mechanisms act to break up the large scale gas structure and inject energy into the ISM, creating more clouds. Ionization creates twice as many clouds compared to the control run without feedback. These clouds are denser and have higher velocity dispersions. Stellar winds only produce 10 per cent more clouds compared to the control run. 
    \item Related to point (iii), each feedback mechanism produces more low-mass clusters of sinks as detected through the DBSCAN algorithm ($<$\SI{e4}{\msol}), especially when both mechanisms are combined. Again, feedback produces smaller gas clouds and thus smaller gas reservoirs for clusters to form from, and feedback reduces accretion onto sink particles resulting in lower mass sinks.
\end{enumerate}
These results show that photoionization is the more important pre-SN mechanism affecting gas dynamics. The main impact of stellar winds is on the sink and cluster properties, due to the formation of cavities around sink particles. However, this requires more investigation, as our sink particles represent collections of many stars. We intend to improve this sub-grid method to resolve individual stars more finely, which will provide more accurate cluster properties. 

\citet{dale2013,dale2014} investigated the effects of stellar winds and photoionization on cloud scales, using similar implementations of feedback as this paper. They found that stellar winds only played a minor role shaping the morphology of GMCs by creating small (of the order of \SI{10}{pc}) bubbles, while photoionization was able to penetrate further and disrupt significant cloud material. They also found that winds were able to disperse dense gas next to stars, and were much less effective at triggering star formation than ionization. Our results agree with these findings on a larger scale, following hundreds of clouds which evolve side by side. However, unlike the Dale et al. simulations, photoionization in our models increases the star formation efficiency rather than decreases it (see also \citealt{bending2020} who discuss this finding further), while stellar winds do not affect it by any significant amount. \citet{grudic2021} investigated larger individual clouds than \citeauthor{dale2014}, ranging in mass from \SI{e6}{} to \SI{e8}{\msol}. Their models showed that radiation (ionizing and as well as non-ionizing radiation) decreased the overall SFE, which is again contrary to our result. However, stellar winds had a negligible impact on the SFE, which we also found.

\citet{gatto2017} found that stellar winds did noticeably decrease the SFE over time-scales of several 10s of Myr. They modelled a vertical slice of a galactic disc ($500\times500\times\SI{5000}{\pc}$) in which smaller SFEs and SFRs were found when winds were included. Both \citeauthor{gatto2017} and our models show more low-mass clusters with winds enabled, showing how this affects the distribution of star formation. Similar models to \citeauthor{gatto2017} were carried out by \citet{rathjen2021}, who found that including ionizing radiation also reduced accretion and formed lower mass clusters. Comparisons between these sets of models must also take into account the different sink properties -- for example, our sinks have accretion radii of \SI{0.78}{\pc} compared to \SI{15.6}{\pc} and \SI{11.7}{\pc}, respectively, and therefore more closely represent stellar (sub)clusters. 

We insert winds in the momentum-conserving phase, unlike other studies which include the energy-conserving phase of the wind as well. This may underestimate the total impact of winds, as an adiabatic wind bubble expands as $R \propto t^{0.6}$ instead of $R \propto t^{0.5}$ \citep{capriotti2001}. \citet{geen2021} injected a hot wind of the form described by \citet{weaver1977}, with the addition of cooling mechanisms; they found efficient cooling at the interface between the wind and ionized gas, effectively rendering the wind expansion as momentum-driven (similar to \citealt{lancaster2021a}). Their models also concluded that winds had a limited influence in disrupting individual molecular clouds. 
Thus even when winds are injected from the energy-conserving phase, their final impact is still secondary to photoionization. However, the shape of the wind bubbles found by \citeauthor{geen2021} are more complex, with plumes or fingers of hot gas expanding preferentially through low density and being cut off by regions of high density. In contrast, our wind bubbles are more spherical, as are \citeauthor{dale2014}'s. Another limitation of our wind method is that we use one main-sequence mass-loss rate per mass bin, neglecting the wind properties of evolved stars. For example, the higher mass-loss rates from Wolf-Rayet stars can lead to faster gas expulsion in embedded clusters \citep[see e.g.][]{rogers2013}.

We do not include other forms of feedback such as radiation pressure. This may play a similar role as stellar winds in forming small-scale cavities around stars, with photoionization still being the dominant mechanism shaping the overall cloud structure \citep{ali2021}. Observationally, the relative impact of radiation pressure is also uncertain, with some studies showing it to dominate over winds or ionization \citep[e.g.][]{olivier2021}, while others show it to be negligible \citep[e.g.][]{mcleod2021}. We also neglect magnetic fields, which may aid winds in driving turbulence on pc-scales \citep{offner2018}. Further limitations of our models, e.g. with regards to photoionization, resolution, initial conditions, and cluster-sinks, have been explored by \citet{bending2020}.

Additionally, we leave SN feedback for future studies. Simulations by \citet{lucas2020} show that the energy from SNe is able to escape clouds through low-density channels created by pre-SN feedback; this energy may have an impact on larger scales beyond an individual cloud. This is supported by some simulations of dwarf galaxies, in which early radiative feedback can aid SNe in driving stronger galactic outflows \citep{hu2017,emerick2018,emerick2019}, while others find weaker outflows due to photoionization producing fewer clusters of SNe \citep{smith2021}.

\section*{Acknowledgements}
We thank the referee for providing helpful comments which improved this paper. We acknowledge funding from the European Research Council for the Horizon 2020 ERC consolidator grant project ICYBOB, grant number 818940. This work was performed using the DiRAC Data Intensive service at Leicester, operated by the University of Leicester IT Services, which forms part of the STFC DiRAC HPC Facility (www.dirac.ac.uk). The equipment was funded by BEIS capital funding via STFC capital grants ST/K000373/1 and ST/R002363/1 and STFC DiRAC Operations grant ST/R001014/1. DiRAC is part of the National e-Infrastructure. Figures were produced using \textsc{splash} \citep{price2007splash}, NumPy \citep{numpy}, Matplotlib \citep{matplotlib}, and Pandas \citep{pandas}.

\section*{Data availability}
The data underlying this paper will be shared on reasonable request to the corresponding author.



\bibliographystyle{mnras}
\bibliography{refs}






\bsp	
\label{lastpage}
\end{document}